\documentclass[a4paper,11pt]{article}
\usepackage{pos}

\title{Integrable Models From Non-Commutative Geometry With Applications to 3D Dualities}
\ShortTitle{Integrable models and 3d dualities}

\author*[a]{Evgeny Skvortsov}
\author[b]{Alexey Sharapov}

\affiliation[a]{Service de Physique de l’Univers, Champs et Gravitation, Universit\'e de Mons, 20 place du Parc, 7000 Mons, Belgium}

\affiliation[b]{Department of Quantum Field Theory, Tomsk State University,\\
Lenin ave. 36, 634050 Tomsk, Russia}

\emailAdd{evgeny.skvortsov@umons.ac.be}
\emailAdd{sharapov@phys.tsu.ru}

\abstract{We discuss a new class of strong homotopy algebras constructed via inner deformations. Such deformations have a number of remarkable properties. In the simplest case, every one-parameter family of associative algebras leads to an $L_\infty$-algebra that can be used to construct a classical integrable model. Another application of this class of $L_\infty$-algebras is related with the three-dimensional bosonization duality in Chern--Simons vector models, where it implements the idea of the slightly-broken higher spin symmetry. One large class of associative algebras originates from Deformation Quantization of Poisson Manifolds. Applications to the $3d$-bosonization duality require, however, an extension to deformation quantization of Poisson Orbifolds, which is an open problem. The $3d$-bosonization duality can be proven by showing that there is a unique class of invariants of the $L_\infty$-algebra that can serve as correlation functions. }

\FullConference{%
  Corfu Summer Institute 2021 "School and Workshops on Elementary Particle Physics and Gravity"\\
  29 August - 9 October 2021\\
  Corfu, Greece
}

\usepackage{tikz-cd,slashed}

\newcommand{\pl}{\partial}

\newcommand{\aA}{{\ensuremath{\mathcal{A}}}}
\newcommand{\aB}{{\ensuremath{\mathcal{B}}}}

\newcommand{\ass}{{\ensuremath{\mathrm{A}_0}}}
\newcommand{\assdef}{{\ensuremath{\mathrm{A}_u}}}

\newcommand{\lA}{{\ensuremath{\boldsymbol{A}}}}
\newcommand{\lL}{{\ensuremath{\boldsymbol{L}}}}
\newcommand{\lgg}{{\ensuremath{\boldsymbol{g}}}}

\usepackage{xcolor}  
\definecolor{myblack}{RGB}{0,0,0}
\definecolor{mygreen}{RGB}{44,105,17}
\definecolor{myblue}{RGB}{34,31,217}
\definecolor{mybrown}{RGB}{194,164,113}
\definecolor{myred}{RGB}{255,66,56}
\definecolor{mymagenta}{RGB}{255,0,255}
\definecolor{mycyan}{RGB}{128,0,0}

\newcommand*{\myblue}[1]{\textcolor{myblue}{#1}}

\newcommand*{\myred}[1]{\textcolor{myred}{#1}}

\newcommand{\stressBlue}[1]{\myblue{\textbf{#1}}}
\newcommand{\stressRed}[1]{\myred{\textbf{#1}}}

\newcommand{\hs}{{\mathfrak{hs}}}

\newcommand{\wv}{{\mathbb{J}}}

\newcommand{\trA}{{\mathrm{Tr}}}

\newcommand{\Trace}[2]{{\mathrm{Tr}_{#1}\!\left[ #2 \right]}}

\newcommand{\Tr}{{\mathrm{Tr}}}

\newcommand{\Trdef}{{\mathrm{Tr}_u}}

\begin{document}
\maketitle
\section{Introduction}
We describe two closely related applications of a new rich class of strong homotopy algebras: (i) integrable models \cite{Sharapov:2019vyd} and (ii) three-dimensional bosonization duality \cite{Sharapov:2018kjz,Gerasimenko:2021sxj}. A cornerstone of these applications is a new approach to constructing strong homotopy algebras  \cite{Sharapov:2018kjz,Sharapov:2018ioy} via intrinsic deformations. On the physics side, the $3d$-bosonization duality conjecture \cite{Giombi:2011kc, Maldacena:2012sf, Aharony:2012nh,Aharony:2015mjs,Karch:2016sxi,Seiberg:2016gmd} takes place in (Chern--Simons) vector models that describe various second-order phase transitions in $3d$, i.e., in the `real physical world'. The class of $L_\infty$-algebras, to be described below, allows one to give rigorous mathematical grounds to the idea of the slightly-broken higher spin symmetry \cite{Sharapov:2018ioy,Sharapov:2018xxx}. Correlation functions are then invariants of this symmetry. At least in the large-$N$ limit the bosonization duality can be reduced to the proof of uniqueness of these invariants. 

Strong homotopy algebras, $A_\infty$, $L_\infty$, $G_\infty$, ..., help to implement the idea of the (most general) consistent algebraic structure. They play a central role in string field theory and provide another point of view on the BV--BRST quantization method, see e.g. \cite{Zwiebach:1992ie,Gaberdiel:1997ia,Kajiura:2003ax,Lada:1992wc,Alexandrov:1995kv,Barnich:2004cr,Hohm:2017pnh,Jurco:2018sby}. However, strong homotopy algebras did not seem to have appeared as symmetries of physical systems in the past. In \cite{Sharapov:2018kjz} it was proposed to implement the idea of the slightly-broken higher spin symmetry \cite{Maldacena:2012sf} as a certain $L_\infty$-algebra. The usual symmetries are realized by Lie algebras whose generators act on one-particle states and the action on multiparticle states is the sum of the actions on each of the one-particle states. There are examples of symmetries that go beyond the standard Lie one. For instance, Yangian is the algebra of conserved charges of an integrable model, whose action on multiparticle states is defined in accordance with a nontrivial co-product and differs from the canonical one.  

Slightly-broken higher spin symmetry extends the notion of symmetries in a different way. First of all, higher spin symmetry is an infinite-dimensional symmetry present in free theories (or in the $N=\infty$ limit of vector models). It is manifested by infinitely many conserved tensors $J_s=j_{a_1\cdots a_s}$ that include the stress-tensor (and global symmetry currents, if present). Canonically, the conserved tensors lead to conserved currents parameterized by Killing tensors and then to conserved charges $Q_s$. The conserved charges $Q_s$ form an infinite-dimensional Lie algebra $\hs$, which includes and extends the usual space-time symmetries. For example, the higher spin currents $J_s$ form a single multiplet of $\hs$, $[Q,J]=J$. It is significant that the Lie algebra $\hs$ 
originates from an associative algebra,\footnote{Symmetries of free/linear equations can be multiplied and, hence, form an associative algebra. The commutator of two is, of course, a symmetry, but the associative structure is more rigid and powerful.} which we denote by the same symbol $\hs$. 
The associative algebra $\hs$ carries all information about the free theory, so that one can write a symbolic equality 
\begin{align*}
    \textbf{Free CFT (QFT) $=$ Associative algebra}\,.
\end{align*}
It is worth noting that the algebra $\hs$ is much smaller than the full operator algebra of a given theory, otherwise the equality above would be rather a trivial statement. Since each higher spin algebra enjoys a canonical trace, $\Tr: \hs\rightarrow \mathbb{C}$, the  correlation functions of the corresponding free theory can be understood as simple trace-type invariants. This can be summarized by the following sequence: 
\begin{align*}
    &\text{Conserved tensor}  &&\rightarrow&& \text{current} &&\rightarrow && \text{symmetry}  &&\rightarrow && \text{invariants$=$correlators} \,.   
\end{align*}
When we turn on interactions or depart from the $N=\infty$ limit, the conservation of higher spin currents is violated ($\pl \cdot J_s\neq 0$) the stress-tensor (and global symmetry currents, if any) remain conserved though. Now, everything depends on the spectrum of primary operators. Vector models have a relatively sparse spectrum of primary operators, as different from models with matter in matrix representations of gauge groups. Therefore, there are  few operators that can appear in the right-hand side of the equality $\pl \cdot J_s=...$. In the large-$N$ limit, it is easy to see \cite{Giombi:2011kc,Maldacena:2012sf} that one can have composite operators of type $[JJ]$. Therefore, the conservation law gets broken by the higher spin currents themselves:
\begin{align}
    \pl \cdot J&= \tfrac{1}{N} [JJ] \label{noncon}\,.
\end{align}
Had we found completely different primary operators on the right-hand side, almost no information could have been extracted. Now, it is possible to close the loop! 

The apparent simplicity of equation \eqref{noncon} is deceptive. Since the currents are not conserved, the charges are no longer conserved as well. As a consequence, the charges do not have to form a Lie algebra anymore, while the higher spin currents may not define a representation of $Q$, i.e.,  
$$[Q,Q]=Q+\ldots \quad \mbox{and}\quad [Q,J]=J+\ldots\,,$$
where the dots stand for $1/N$ corrections. One can also show that the algebra $\hs$ is rigid  and the currents $J$, as an $\hs$-module, cannot be deformed as well. Hence, it is impossible to account for the dots on the right by any usual deformation of a Lie algebra or/and its representation.  The main proposal made in \cite{Sharapov:2018kjz} is that we should deform the whole structure ``Lie algebra plus its module'' into a strong homotopy Lie algebra. Now, the lowest structure maps $l_2(\bullet,\bullet)$ simply encode $\hs$ and $J$ as its module. Higher structure maps $l_n(\bullet,\ldots,\bullet)$ allow one to deform the Lie algebra $\hs$ together with its module $J$, so that they form a single algebraic structure. 

To summarize, the new type of symmetry we propose deforms a Lie algebra and module(s) of this algebra into a single $L_\infty$-algebra. Such a structure requires the $L_\infty$ to be made of two components as a graded vector space: one for the initial Lie algebra and another for its module(s). Therefore, this can also be understood as a Lie algebroid, but we will not exploit this interpretation. Now, the problem of correlation functions is equivalent to the problem of invariants of this $L_\infty$-algebra. 

In  applications to the $3d$-bosonization duality conjecture, $\hs$ is the Weyl algebra, i.e., the universal enveloping of  the Lie algebra of canonical commutation relations $[q^i,p_j]=i\hbar \delta^i_j$. It results from the  deformation quantization of the simplest Poisson Manifold. The $L_\infty$ algebra is completely determined by a closely related algebra, which can be understood as a deformation quantization of the simplest Poisson Orbifold $\mathbb{R}^2/\mathbb{Z}_2$. The last fact implies a deep relation between the two problems. For the simplest case of $O(N)$ vector models, one can see \cite{Sharapov:2020quq} that the $L_\infty$-algebra depends on two phenomenological parameters, which can be related with  $1/N$ and the Chern--Simons level $k$. It is also possible 
to show \cite{Sharapov:2020quq} that the trace invariants of $\hs$ are unique invariants that deform to the invariants of the $L_\infty$-algebra. The $3d$-bosonization duality is a simple consequence of the uniqueness: since the correlation functions are completely specified by the symmetry, it does not matter if the underlying microscopical realization involves bosonic or fermionic degrees of freedom.  

It should be remembered that  \eqref{noncon} is an exact quantum equation of motion, which is true even for $N=1$ (the case of the Ising model). It is unclear at the moment if the semi-classical arguments based on $L_\infty$ can be extended to the quantum domain. What makes them semi-classical 
is a  simple form of  relation between $J$ and $[JJ]$ suggested by the representation theory. These operators, however, renormalize in a nontrivial way. Therefore, it would be important to understand if `quantization' of the $L_\infty$-symmetry is possible to be able to extend the proof beyond the large-$N$ limit.

Let us give an overview of other results that rely on the same construction \cite{Sharapov:2018kjz,Sharapov:2018ioy} of $L_\infty$-algebras. In its simplest realization the idea of intrinsic deformations allows us to construct an $A_\infty$-algebra from any associative algebra $A_u$ that depends on a parameter $u$. Canonically, every $A_\infty$-algebra generates an $L_\infty$-algebra via the anti-symmetrization map. On the other hand, every $L_\infty$-algebra can be identified with a homological vector field $Q$ on a formal graded manifold $\mathcal{N}$ with coordinates  $\Phi^\aA$.  Given such a homological vector field $Q$, we can construct a sigma-model with the following equations of motion:
\begin{align}
    d\Phi&= Q(\Phi)\equiv l_2(\Phi,\Phi)+l_3(\Phi,\Phi,\Phi)+\ldots\,.\label{gensigma}
\end{align}
Here the fields $\Phi^{\aA}$ define a map from the odd tangent bundle $T^\ast[1]\mathcal{M}$ of a space-time manifold $\mathcal{M}$ to the target space $\mathcal{N}$. This construction can, in principle, be applied to any field theory, as one can always formulate classical equations of motion in the form  \eqref{gensigma} and there is a canonical procedure of assigning a homological vector field $Q$ to any field theory \cite{Barnich:2004cr,Barnich:2010sw}.  

The class of strong homotopy algebras obtained via intrinsic deformations has a number of special properties. First and foremost, all the multilinear  maps $l_n$ entering the r.h.s. of (\ref{gensigma}) can be constructed from the bilinear maps  
\begin{align*}
     a\circ b= a\,\star\, b+\sum_{k=1}\phi_k(a,b)\,u^k
\end{align*}
that determine expansion of the associative product $\circ$ in $A_u$ around ${u=0}$. Second,  the resulting classical field theory/mechanics appears to be integrable, e.g., one can construct all its solutions with the help of a Lax pair. To summarize, each one-parameter family of associative algebras gives rise to a classical integrable model:
    \begin{align*}
        \textbf{{Soft${}_{u}$ Associative Algebras $\longrightarrow$ Integrable models}}\,.
    \end{align*}
As an interlude we also discuss Deformation Quantization of Poisson Orbifolds as a rich source of one-parameter families of algebras; in particular, we discuss $\mathbb{R}^2/\mathbb{Z}_2$, which is relevant to the $3d$-bosonization duality. At present, this is largely an open problem since no analog of Kontsevich's formality is known for the case of Orbifolds. 

\section{Integrable models from associative algebras}
\label{sec:int}
A Lax pair is a cornerstone of integrable models. As the name suggests, it is a pair of two matrices, usually called $L$ and $A$, that depend on the `time' $t$ and, possibly, on other parameters. The matrices  are required to satisfy the Lax equation
\begin{align}
    \pl_t L&=[A,L]\,. 
\end{align}
Often, $A$ is a given function of $L$, so this is a nonlinear equation for $L$ as a function of $t$. 
As a result, one gets immediately an infinite family of integrals of motion $I_n=\Tr[L^n]$, $n=1,2,\ldots$, not all of which may be independent. Clearly, $A$ plays the role of connection along $t$,  and $L$ is required to be covariantly constant. This suggests a natural generalization of the Lax pair to $d$ dimensions. Extension to $d$ dimensions is not unique as one can have more structures. Given a $d$-dimensional manifold $\mathcal{M}_d$, introduce a connection $A$ on some vector bundle over $\mathcal{M}_d$ where a zero-form $L$ takes its values and set
\begin{align}
    dA&=\tfrac12[A,A]\,,& dL&=\rho(A)L \label{laxd}\,.
\end{align}
Here the covariant derivative is (implicitly) $D_A=d-\rho(A)$, where $\rho$ denotes the action of $A$ on $L$. Optionally, one can introduce forms of various degrees that are constrained to be covariantly constant with respect to $A$. Assume further that  both $A$ and $L$ take values in the algebra $A_0=\mathrm{End}(V)$ of endomorphism of a vector space $V$. Such $A_0$, depending on a situation, can be viewed as an associative algebra or as a Lie algebra with the Lie bracket defined by the commutator. 
Setting $\rho(A)L=[A,L]$ and using the trace on $
A_0$, we obtain an infinite sequence of integrals of motion  $I_n=\Tr[L^n]$. It may also be convenient sometimes to solve the generalized Lax's equations (\ref{laxd}) in a `pure gauge' form:
\begin{align}\label{formalso}
    A&= g^{-1} dg \,,& L&=g^{-1}L_0g\,,
\end{align}
where $L_0$ is constant, $dL_0=0$. The zero-form $L_0$ parameterizes the solution space locally, but not necessarily globally. The simplest Lax pair \eqref{laxd} will provide integrability of the class of models to be introduced shortly.

\paragraph{Interactions.} A model that can explicitly be reduced to \eqref{laxd} can be thought of as non-interacting one, which is justified by local solutions \eqref{formalso}. We would like to study the most general deformation of \eqref{laxd}. Based on the simple form-degree counting, a generic deformation is of the form
\begin{equation}\label{mostgeneral}
    \begin{array}{rcl}
         dA&=&l_2(A,A)+l_3(A,A,L)+l_4(A,A,L,L)+\ldots=F_A(A;L)\,,\\[3mm]
    dL&=&l_2(A,L)+l_3(A,L,L)+\ldots=F_L(A;L)\,, 
    \end{array}
\end{equation}
where $l_n$ are various interaction vertices, or structure maps, with all arguments indicated explicitly. The formal sum thereof gives two structure functions $F_A$ and $F_L$. The bilinear maps are known from \eqref{laxd} to define a Lie algebra (where $A$ takes the values) and its module  (where $L$ takes the values). In our case, the algebra is $A_0=\mathrm{End}(V)$ for some vector space $V$ and the module is the adjoint one. 

Clearly, \eqref{mostgeneral} is a particular case of a more general set up where one has a space-time manifold $\mathcal{M}_d$ and a formal graded manifold $\mathcal{N}$ equipped with a homological vector field $Q$. If $\Phi=\{ \Phi^\aA\}$ are coordinates on $\mathcal{N}$, then $Q=Q^\aA \pl/\pl \Phi^\aA$ and the defining condition for an odd vector field $Q$ to be homological reads\footnote{The derivative with respect to $\Phi^\aA$ is understood in the appropriate graded sense. }
\begin{align}
    Q^2&=0 &&\Longleftrightarrow && Q^\aB \frac{\pl}{\pl \Phi^\aB} Q^\aA=0\,.
\end{align}
If $\mathcal{N}$ is non-negatively graded, one can write down a sigma-model by promoting $\Phi^\aA$ to smooth maps $\Phi^\aA: T^\ast[1]\mathcal{M}_d\rightarrow \mathcal{N}$ of degree zero. Geometrically, on can think of these maps  as differential forms $\Phi^\aA(x, dx)$ on $\mathcal{M}_d$ with values in $\mathcal{N}$. The sigma-model\footnote{As a historical note, systems of the form \eqref{gensigma} were first introduced by Sullivan \cite{Sullivan77} as Free Differential Algebras, re-introduced into physics \cite{vanNieuwenhuizen:1982zf,DAuria:1980cmy} in the supergravity and higher spin \cite{Vasiliev:1988sa} contexts. }
\begin{align}\label{mevenmore}
    d \Phi&= Q(\Phi) 
\end{align}
is defined by requiring the map to respect the differentials, which is exactly \eqref{mevenmore}. The most common situation is when $\Phi$ are coordinates at a vicinity of a stationary point, i.e., $Q(0)=0$. The Taylor expansion of $Q$ defines then multilinear graded-symmetric maps $l_n(\bullet,\ldots,\bullet)$ that satisfy Stasheff's relations for an $L_\infty$-algebra. Eq. \eqref{mevenmore}, regarded as a set of PDE's for the form fields $\Phi^\aA(x, dx)$, has a very important property, which can be phrased in several equivalent ways: (i) there are no any hidden algebraic constraints  on the fields buried in \eqref{mevenmore}; (ii) Eq. \eqref{mevenmore} is consistent with $dd\equiv 0$, which is the Frobenius integrability condition. Sadly, this is not a type of integrability that helps to solve the model.

From the modern point of view, every gauge PDE defines and is defined by a $Q$-manifold $\mathcal{N}$, see e.g. \cite{Grigoriev:2019ojp}. In a few words, given a canonical BV-BRST formulation of a gauge theory, one can consider its jet space extension \cite{Barnich:2004cr,Barnich:2010sw,Grigoriev:2019ojp}, which is an $L_\infty$-algebra. This algebra may have positive and negative degrees corresponding to ghosts (ghosts for ghosts etc.) as well as to anti-fields. A useful `derivative' of any $L_\infty$-algebra is its {\it minimal model}, which, being much smaller, contains all the essential information (it is said to be quasi-isomorphic to the initial $L_\infty$-algebra). For a large class of models, e.g. field theories, such minimal model is a non-negatively graded $L_\infty$-algebra. It is the corresponding $Q$, $Q^2=0$, that can be thought of as an invariant definition of the field theory we started with \cite{Grigoriev:2019ojp}. At the same time $Q$ is in the possession of all local relevant information about the field theory, e.g. \eqref{mevenmore} defines solutions of the classical field equations. 

As a result, we have system \eqref{mostgeneral} that is a particular case of \eqref{mevenmore}. Here, $\Phi=\{A, L\}$ and $Q=F^A \pl/\pl A+ F^L \pl/\pl L$. It is a smooth deformation of a trivial one, which is the Lax pair. The deformation is required to be nontrivial: (i) in the field theory language, there is no field redefinition that brings \eqref{mostgeneral} into \eqref{laxd}; (ii) in the $L_\infty$-language, $l_n(\bullet,\ldots,\bullet)$, $n>2$ cannot be eliminated by a natural transformation of $L_\infty$-algebras; (iii) in the $Q$-manifold language, there is no coordinates in which the homological vector field assumes the quadratic form, $Q=\Phi\Phi \pl/\pl \Phi$.   

System \eqref{mostgeneral} is yet to be constructed! We arrive at a number of questions regarding \eqref{mostgeneral}: (a) how to construct vertices $l_n$ given the initial data encoded in $l_2$? (b) how to solve it? Under certain assumptions \cite{Sharapov:2017yde,Sharapov:2018kjz,Sharapov:2019vyd} it can be shown that the $L_\infty$ originates from a certain $A_\infty$ and the latter is built from a deformation of $A_0$ as an associative algebra. Solutions can also be constructed explicitly. 

We recall that $l_2$ originates from an associative product on $A_0$. Let us first recast these initial data into $L_\infty$-language. In accordance with the form degrees, our $\mathcal{L}$ consists, as a vector space, of two copies of $A_0$ that are assigned degrees $0$ and $1$, i.e. $\mathcal{L}=\mathcal{L}_0\oplus \mathcal{L}_1$, $\mathcal{L}_{0,1}\sim A_0$. The bilinear structure maps are defined as
\begin{align}
   l_2(a,b)&=[a, b]_\star \,, & l_2(a,v)&=a\star v-v\star a \,, &&  a,b\in \mathcal{L}_1\,, \quad  v\in \mathcal{L}_{0}\,.
\end{align}

In fact, the associative structure is of immense help. The underlying structure will turn out to be $A_\infty$ rather than $L_\infty$.\footnote{For completeness, $A_\infty$ is a graded vector space, equipped with degree $(-1)$ maps $m_n(\bullet,\ldots,\bullet)$ that satisfy $m\circ m=0$, where $m=m_1+m_2+...$ is a formal sum and $\circ$ is the Gerstenhaber product. The latter is defined for any two multilinear maps $f$ and $g$ of degrees $|f|$ and $|g|$ and having $k_f$ and $k_g$ arguments as
\begin{align}
    f\circ g&= \sum_i (-1)^\kappa f(a_1,\ldots,a_i, g(a_{i+1},\ldots,a_{i+k_g}),a_{i+k_g+1},\ldots, a_{k_f+k_g-1})   \,.
\end{align}
Here $\kappa$ is the usual Koszul sign: $\kappa=|g|(|a_1|+\cdots+|a_i|)$. Similar relations define $L_\infty$-algebras. However, it is more handy to get them from the Taylor expansion of $Q^2=0$. 
} Anticipating this fact, the same initial data can be encoded by a small $A_\infty$-algebra:
\begin{align}\label{smallA}
   m_2(a,b)&=a\star b \,, & m_2(a,v)&=a\star \nu\,, & m_2(v,a)&=-v\star a \,, && a,b\in \mathcal{A}_1\,, \quad v\in \mathcal{A}_{0}\,.
\end{align}
Our $L_\infty$-algebra will come from an $A_\infty$-algebra via the usual anti-symmetrization map
\begin{align}\label{asym}
    l_n(x_1, \ldots, x_n)=\sum_{\sigma\in S_n} (-1)^{|\sigma_x|} m_n(x_{\sigma(1)},\ldots,x_{\sigma(n)})\,.
\end{align}
We would like to deform \eqref{smallA} as to activate higher structure maps $m_n$, $n>2$, in a nontrivial way.

\paragraph{Main result.} With all definitions above we can now formulate the main result \cite{Sharapov:2018kjz,Sharapov:2019vyd}. We need to deform the $A_\infty$-algebra \eqref{smallA} that consists of an associative algebra $\ass$ together with its natural bi-module. Under some technical assumptions\footnote{The main assumption is that we are looking for a deformation that makes sense when $A_0$ is replaced by $\mathrm{Mat}_N(A_0)$ for any $N$. Obviously, \eqref{smallA} can be defined for $\mathrm{Mat}_N(A_0)$. It is this assumption that allows us to reduce a complicated problem of Chevalley--Eilenberg cohomology to a much simpler Hochschild cohomology. Nevertheless, in some examples of practical importance the $N=1$ case seems to be as general as any $N>1$.}, the deformation exists iff the underlying associative algebra is soft, i.e. can be deformed into a (nontrivial, of course) one-parameter family of associative algebras $\assdef$. Therefore, the product in $\assdef$ deforms the one in $\ass$:
\begin{align}\label{defproduct}
     a\circ b= a\,\star\, b+\sum_{k=1}^\infty\phi_k(a,b)\,u^k\,.
\end{align}
That the deformed product is associative, $a\circ (b\circ c)= (a\circ b)\circ c$, imposes certain conditions on the bilinear maps $\phi_k$. The nontriviality of the deformations implies that $\phi_1$ is a nontrivial class of the second Hochschild cohomology group $HH^2(\ass,\ass)$.

The algorithm for constructing the model is as follows. First, one constructs a deformation of $\mathcal{A}_0$ as of an $A_\infty$-algebra. The first few structure maps read
\begin{align*}
    m_3( a, b, v)&=\phi_1( a, b)\,\star\, v\quad \rightarrow\quad l_3\,,\\
    m_4( a, b, v, w)&=\phi_2( a, b)\,\star\, v\,\star\,w +\phi_1(\phi_1( a, b),v)\,\star\, w \quad  \rightarrow \quad l_4\,.
\end{align*}
There is a number of ways \cite{Sharapov:2018kjz} to get all $m_n$, including an explicit formula for any $n$. Secondly, one induces the corresponding $L_\infty$-maps by means of  anti-symmetrization (\ref{asym}). As a result we have all $l_n$ or $Q$ and it can be shown that model \eqref{mostgeneral} so defined is not equivalent to a free one via field-redefinitions.

Nevertheless, we can show that the model is integrable and its solution space can be understood via an auxiliary Lax pair \cite{Sharapov:2019vyd}. Not surprisingly the Lax pair, which has the same form as \eqref{laxd}, is based on the deformed product $\circ$, namely, 
\begin{align}
    d\lA&=\lA\circ\lA \,, & d\lL&=\lA\circ\lL-\lL\circ\lA \,,\label{laxdAux}
\end{align}
where $\lA$, $\lL$ take values in the associative algebra $\assdef$. The integrability means here that solutions of \eqref{mostgeneral} can be built from those of \eqref{laxdAux}, the latter having pure gauge form
\begin{align}\label{formalsol}
    \lA&= \lgg^{-1}\circ d\lgg \,,& \lL&=\lgg^{-1}\circ\lL_0\circ\lgg\,.
\end{align}
The map is constructed order by order in $u$ as\footnote{Similar maps relating interacting and quasi-free or non-commutative and commutative theories are known in the literature \cite{Nicolai:1980jc,Prokushkin:1998bq,Seiberg:1999vs}.}
\begin{align}
    A&= \lA +(\pl_u \lA) \circ \lL +\ldots \Big|_{u=0}\,, & L&= \lL+(\pl_u \lL)\circ \lL +\ldots  \Big|_{u=0}\,.
\end{align}
Provided the associative algebra $\assdef$ admits a trace $\Trdef$, that is, $\Trdef[a\circ b-b\circ a]=0$, we can also define the integrals of motion $I_n=\Trdef[\lL\circ \cdots \circ \lL]$ obeying the conservation law $dI_n=0$.\footnote{Note that this invariants differ from what is usually called integrals of motion (function(al)s of fields that are time independent). In modern terms, integrals of motion correspond to zero-form symmetries and are given by closed $(d-1)$-forms. Due to $dI_n=0$ functions $I_n$ do not depend on space-time points at all and correspond to $(d-1)$-form symmetries, \cite{Gaiotto:2014kfa}.} For infinite-dimensional algebras there can be other types of integrals of motion \cite{Sharapov:2020quq}.

To summarize, given a one-parameter family of associative algebras $\assdef$, one can construct a non-linear system of formally consistent\footnote{A word of warning is that formal consistency does not imply actual consistency. It is easy to construct examples of formally consistent equations that do not make any sense, i.e. they are just symbolic expressions, but lead to infinite tree-level amplitudes. See e.g. \cite{Boulanger:2015ova,Skvortsov:2015lja} for explicit  examples. } PDE's that, nevertheless, can be solved via an auxiliary Lax pair.

\section{Deformation quantization of Poisson Orbifolds.} 
\label{sec:DQ}
A natural question is where to find either soft associative algebras (to be deformed) or one-parameter families of associative algebras right away. There are two large stocks of such algebras: (1) finite dimensional associative algebras; (2) deformation quantization. 

All finite-dimensional semi-simple associative algebras, by the Artin--Wedderburn theorem, are products of matrix algebras. As such they cannot have free parameters. Therefore, soft algebras have to be outside the nice class of semi-simple algebras. The classification of the latter is known to be wild. Good news: there are a lot of soft algebras along these lines. Bad news: it is hard to find them. Below we discuss the item (2) in more detail as it is of some interest due to its relation to the $3d$ bosonization duality. 

Deformation quantization is an obvious source of one-parameter associative algebras. Let us recall the main idea. Given any Poisson manifold $\mathcal{P}$ one considers the algebra of functions $C^\infty(\mathcal{P})$. It is an associative and commutative algebra. One then tries to deform the product
\begin{align}
    f\star g&= f\cdot g +\hbar \{f,g\}+\mathcal{O}(\hbar^2)\,,
\end{align}
where the first order deformation is the Poisson bracket on $\mathcal{P}$, which, at same time, is a nontrivial Hochschild two-cocycle of $C^\infty(\mathcal{P})$. The general solution of this problem was given by Kontsevich in \cite{Kontsevich:1997vb}. Later \cite{Cattaneo:1999fm}, it was shown that Kontsevich's graphs can be understood as Feynman's diagrams of an auxiliary topological string theory, called Poisson sigma-model. Deformation quantization gives a large class of one-parameter families of associative algebras, parameterized by Poisson manifolds. 

The space of algebras can be extended even further, leading to an interesting  open problem in deformation quantization. Poisson manifolds generally enjoy a lot of  symmetries. For a given Poisson manifold $\mathcal{P}$ one can pick up some group of discrete symmetries $\Gamma$. There are two associative algebras naturally linked to these data: 
\begin{enumerate}
    \item $\Gamma$-invariant functions $C^\infty(\mathcal{P})^\Gamma\sim C^\infty(\mathcal{P}/\Gamma)$, where we see the orbifold $\mathcal{P}/\Gamma$. This algebra is commutative;
    \item Cross-product algebra $C^\infty(\mathcal{P}) \ltimes \Gamma$. The algebra is generated by sums $\sum_i f_i \otimes \gamma_i$, where the $f_i\in C^\infty(\mathcal{P})$ and $\gamma_i\in \Gamma$. The product is defined as $(f \otimes \gamma)\diamond (f' \otimes \gamma')=(f \gamma(f'), \gamma\gamma')$, where $\gamma(f)$ denotes the action of  $\gamma\in \Gamma$ on a function $f\in C^\infty(\mathcal{P})$. Note that this algebra is generally  non-commutative
    even for abelian $\Gamma$.
\end{enumerate}
The two algebras above are closely related to each other.  From the viewpoint of non-commutative geometry the second algebra is more favorable. In some sense, it encodes complicated information about singular manifolds like orbifolds through the non-commutativity of the algebra. 

In both cases one can ask: (i) what are nontrivial deformations? (ii) how to construct them to all orders? We will refer to both of the problems, i.e. for algebras (1) and (2) above, as to the problem of Deformation Quantization of Poisson Orbifolds. More generally: is there any counterpart of Kontsevich's Formality Theorem for Poisson Orbifolds? 

The first surprising result is that Poisson Orbifolds have more deformations! There are other directions of deformation quantization that rely on orbifolding (or on the extension by $\Gamma$). These new deformations are not captured by Kontsevich's Formality at all. It is also not clear what kind of Formality one should be looking for \cite{Tiang,Halbout}. In principle, following Kontsevich (who mentioned in \cite{Kontsevich:1997vb} that his construction originates from a topological string theory, which has been then manifested by the quantization of the Poisson sigma-model in \cite{Cattaneo:1999fm}) one should quantize the same Poisson sigma-model on Poisson Orbifolds, which may crucially depend on the type of an orbifold, preventing one from seeking out any universal extension of the Formality Theorem. 

While the general problem of Deformation Quantization of Poisson Orbifolds is open, there is a number of cases, where a constructive approach is possible. For example, let us consider the Weyl algebra $A_n$ extended by a finite group of symplectomorphisms $\Gamma \subset \mathrm{Sp}_n$, i.e. $A_n\ltimes \Gamma$. The Weyl algebra can be understood as a result of the deformation quantization of $\mathbb{R}^{2n}$ endowed with the canonical symplectic form $dp_i\wedge dq^i$. Now, on top of that we can have several new deformations. The number of such deformations depends on $\Gamma$. The deformations can be constructed explicitly to all orders via injective resolutions \cite{Sharapov:2017lxr,Sharapov:2018hnl}. 

Let us consider in more detail the case of the  algebra $A_1\ltimes \mathbb{Z}_2$, as it is relevant for the $3d$ bosonization duality. Here $A_1$ can be understood as an associative algebra generated by the pair of canonical variables subject to the commutation relation
\begin{align}
    [q,p]&=i\hbar\,,
\end{align}
i.e. we have already quantized $\mathbb{R}^2$ along $dq\wedge dp$. The group of symplectic reflections $\Gamma=\mathbb{Z}_2$ is generate by a single element  $R$ that sends $(q,p)$ to $(-q,-p)$. The crossed-product algebra $A_1\ltimes \mathbb{Z}_2$ is obtained by adding the new generator $R$ to $q$ and $p$ such that \begin{align}
    R^2&=1\,, && RqR=-q \,, && RpR=-p\,.
\end{align} 
The general element of $A_1\ltimes \mathbb{Z}_2$ is given by a polynomial $f(q,p,R)=f_0(q,p)+f_1(q,p)R$ and $R$ acts via the reflection whenever we need to drag it through $q$ or $p$. We can also start from scratch: set $\hbar=0$. The algebra $\mathcal{A}_{0,0}$ of functions $f(q,p,R)$ is still non-commutative due to $R$. As an associative algebra $\mathcal{A}_{0,0}$ admits a two-parameter family of deformations, the usual one along the classical Poisson bracket $\{\bullet,\bullet\}$ and another one, along $R$. We can also jump directly to $\mathcal{A}_{\hbar,0}=A_1\ltimes \mathbb{Z}_2$. The latter algebra features the second deformation along $R$.

Independently of the ideas above, the deformation of $\mathcal{A}_{\hbar,0}$ was found by Wigner \cite{Wigner1950} and explored further in a great number of papers, see e.g. \cite{Yang:1951pyq, Boulware1963, Gruber, Mukunda:1980fv}. It is remarkable, that one can generate the deformed algebra $\mathcal{A}_{\hbar,u}$ by deforming the canonical commutation relations \cite{Yang:1951pyq, Boulware1963, Mukunda:1980fv}
\begin{align}
    [q,p]&=i\hbar+iuR\,.
\end{align}
Based on this relation, one can work out the structure constants, see \cite{Pope:1990kc,Bieliavsky:2008mv,Joung:2014qya,Korybut:2014jza,Basile:2016goq,korybut2020star} for different approaches to the problem. There are closely related algebras: the $R$-invariant subalgebra, known as $gl_\lambda$ \cite{Feigin1987}, and the non-commutative hyperboloid \cite{Bieliavsky:2008mv} (or fuzzy sphere, depending on the real form).

It is also interesting that the first order deformation of $\mathcal{A}_{\hbar,0}$ can be derived \cite{Sharapov:2017yde} from an extension of Kontsevich's Formality known as the Shoikhet--Tsygan--Kontsevich formality \cite{Tsygan,Shoikhet:2000gw}. We start from $A_{\hbar,0}$, which can be realized as the algebra of functions $f(q,p,R)$ equipped with the Moyal--Weyl star-product, $f\star g$. Extension with $R$ is easy and requires us to first move all $R$ to the left (or to the right), while properly commuting them with $q,p$, and then take the star-product. The first order deformation along $R$,
\begin{align}
    f\circ g&= f\star g +u\, \phi(f,g) R+\mathcal{O}(u^2)\,,
\end{align}
is defined by a Hochschild two-cocycle $\phi$. It is this two-cocycle that can be computed with the help of Shoikhet-Tsygan-Kontsevich formality. None of the known formality theorems seems to have anything to say about higher orders.

\section{Slightly-broken higher spin symmetry and three-dimensional bosonization duality}
Another application of the $A_\infty$- and $L_\infty$-algebras constructed in section \ref{sec:int} is to the three-dimensional bosonization duality conjecture \cite{Sharapov:2018kjz,Gerasimenko:2021sxj}. What  is interesting here is that $L_\infty$-algebras manifest themselves as symmetries of physical models. We begin with a brief overview of the simplest models conjectured to exhibit the duality. Next, we discuss why and how these models exhibit infinite-dimensional symmetries that are realized as $L_\infty$-algebras. Correlation functions can then be understood as invariants of this symmetry, while the proof of their uniqueness leads to a proof of the duality.

\paragraph{Higher spin symmetry of free theories.} It has long been known that free QFT's (as well as the classical field theories they are based on) feature infinite-dimensional symmetries that are generated by infinitely many conserved tensors on top of the usual low spin stress-tensor (spin-two) and global symmetry currents (spin-one). Such symmetries were dubbed Zilch symmetries \cite{Zilch, Deser:1980fk}. In modern terminology they are called higher spin currents and symmetries. The conserved tensors, e.g. for the free scalar theory, 
\begin{align}
    J_{a_1\cdots a_s}&= \bar \phi \pl_{a_1}\cdots \pl_{a_s} \phi +\ldots 
\end{align}
are sandwiches of two fields with an arbitrary number of derivatives. The omitted terms can be arranged to make it conserved as a consequence of $\square \phi=m^2\phi$. In the momentum space, the existence of the corresponding charges is obvious since any
\begin{align}
    Q_f&= \int d^{d-1}p\, \bar{a}_p f(p, \nabla_p) a_p
\end{align}
is a conserved charge for arbitrary kernel $f$. Locality of the conserved tensors imposes mild restrictions on kernels $f$ \cite{Zilch}. Charges $Q_f$ map one-particle states $V$ to themselves, i.e. they belong to $\mathrm{End}(V)$. For QFT's in Minkowski space $\mathrm{End}(V)$ contains the Poincar\'e algebra and for massless QFT's it usually includes the conformal algebra. As a useful toy model we can think of the massless scalar field, which is conformally-invariant.

A closely related classical concept is that of the (higher) symmetries of field equations, i.e. maps $S$ that map solutions of some equation  $E\phi=0$ to solutions. Such maps have to obey $ES=S'E$ for some $S'$ that depends on $S$. For the free massless scalar the conformal transformation reads
\begin{align}\label{confs}
    \delta_v \phi&= v^m \pl_m \phi +\frac{\Delta}{d} (\pl_m v^m) \phi\,,
\end{align}
where $\Delta=(d-2)/2$ is the conformal weight and $v^m(x)$ is a conformal Killing vector. The stress-tensor $J_{ab}$ gives rise to a number of currents $j_m= J_{mb}v^b$ that are parameterized by Killing vectors. The infinitesimal transformations $\delta_v$ form the conformal algebra $so(d,2)$ under commutators. Likewise, given a conserved rank-$s$ tensor, one can construct a number of conserved currents
\begin{align}
    j_m(v)&= J_{ma_1\cdots a_{s-1}}\, v^{a_1\cdots a_{s-1}}\,, &  \pl^{(a_1} v^{a_2\cdots a_s)}-\text{traces}&=0\,,
\end{align}
where $v^{a_1\cdots a_{s-1}}$ is a conformal Killing tensor.\footnote{As an $so(d,2)$-module the space of solutions for $v^{a_1\cdots a_k}$ can be identified \cite{Eastwood:2002su} with an irreducible tensor $v^{A_1\cdots A_k,B_1\cdots B_k}$, $A,B,\ldots =0,\ldots ,d+1$ being $so(d,2)$-indices. This tensor is symmetric in $A$'s and $B$'s, separately. It is traceless and subject to the Young condition $v^{(A_1\cdots A_k,A_{k+1})B_2\cdots B_k}=0$.} The charges $Q_s(v)$ corresponding to higher spin currents $j_m(v)$ form an infinite-dimensional Lie algebra, denoted by $\hs$, that contains $so(d,2)$ as a subalgebra. 

Thanks to the linearity of the equations of motion, the algebra of symmetries is associative, i.e. the product $S_1S_2$ of any two symmetries $S_{1,2}$ is a symmetry again. One does not have to take the commutator $[S_1,S_2]$ to get a symmetry. In particular, one can multiply a number of the conformal symmetries \eqref{confs} to get a genuine higher spin symmetry. An important conclusion is that the algebra of symmetries is an infinite-dimensional associative algebra, still called $\hs$ and referred to as the higher spin algebra, that contains $so(d,2)$ as a Lie subalgebra. (One can always get a Lie algebra out of an associative one by defining the Lie bracket as the commutator.) Higher spin algebras `know' everything about the underlying free field theories, which we summarize as
$$
\mbox{\bf Free CFT = Associative Algebra}\,.
$$
The arguments above suggest that any higher spin algebra (for a conformally invariant theory) can be understood as $U(so(d,2))/I$ for a suitable two-sided ideal $I$ \cite{Eastwood:2002su}. Different matter systems lead to different higher spin algebras, e.g. the higher spin algebra of the free boson is not isomorphic to that of the free fermion in $d>3$. 

The $3d$ case is very special. First of all, one-particle states of the free boson $|\phi\rangle$ and free fermion $|\psi\rangle$ can be realized, respectively,  as even and odd subspaces of the Fock space $f(\hat{a}^\dag_i)|0\rangle$ of the $2d$ harmonic oscillator $[\hat{a}^i, \hat{a}^\dag_j]=\delta^i_j$, $i,j=1,2$. This is partly thanks to the isomorphism $so(3,2)\sim sp(4,\mathbb{R})$ and the latter algebra  has the standard oscillator realization where the generators of $sp(4,\mathbb{R})$ can be identified with bilinears in the ladder operators. In particular, translations $P_m\sigma^m_{ij}$ act as $\hat{a}^\dag_i\hat{a}^\dag_j$, where $\sigma^m_{ij}$ are the Pauli matrices. Secondly, the higher spin algebra of the free fermion coincides with that of the free boson. Thirdly, they are both isomorphic to the even subalgebra $A_2^e$ of the Weyl algebra $A_2$ \cite{Dirac:1963ta, Gunaydin:1981yq,Gunaydin:1983yj}, i.e. to the algebra of even functions $f(\hat{a},\hat{a}^\dag)=f(-\hat{a},-\hat{a}^\dag)$. The fact that the higher spin algebras of the free boson and fermion are isomorphic is the starting point for the proof of the three-dimensional bosonization duality. 

Correlation functions in a CFT must be conformally-invariant. Likewise, correlators in a CFT with a higher spin symmetry $\hs$ should be invariant under the full $\hs$. Therefore, one can ask if $\hs$ as a Lie algebra admits any invariants that can serve as correlation functions provided one computes them on appropriate elements. Indeed, each higher spin algebra is equipped with an invariant trace $\Tr_\star[\bullet]$, i.e. $\Tr_\star[a\star b-b\star a]=0$, where $\star$ denotes the product in $\hs$. The $\star$-notation is not an accident. In the case of $3d$ the product in the Weyl algebra can be realized as the Moyal--Weyl star-product.\footnote{In general, higher spin algebras can be understood as a result of deformation quantization, see e.g. \cite{Michel,Joung:2014qya} and references therein.} It comes as no surprise that all $n$-point correlation functions of $J$ can be computed at once as higher spin invariants
\begin{align}\label{correlatorsFree}
    \langle J_1\ldots J_n\rangle= \Trace{\star}{\wv_1\star \cdots \star \wv_n}\,,
\end{align}
provided $\wv_i$ are chosen appropriately to behave as quasi-primary operators that represent a generating function $J_i\equiv J(x_i)$ of higher spin currents at point $x_i$. Invariance under the higher spin and, hence, under the conformal symmetry is manifest:
\begin{align}
    Q_s \langle J_1\ldots J_n\rangle&=0 && \Longleftrightarrow && \delta_\xi \Trace{\star}{\wv_1\star \cdots \star \wv_n}=0\,,
\end{align}
where $\delta_\xi \wv=[\xi,\wv]_\star$, $\xi\in \hs$ incorporates the action of the higher spin algebra, including the conformal symmetries, of course. Transformation $\delta_\xi \wv$ represents the action $[Q,J]=J$ of higher spin charges $Q$ onto the higher spin current multiplet. $\delta_\xi$ form the same higher spin algebra $\hs$ as a Lie algebra, which corresponds to $[Q,Q]=Q$ - type relations at the level of charges.

The calculation of the invariants is simple and gives all $n$-point functions \cite{Colombo:2012jx,Didenko:2012tv,Didenko:2013bj, Bonezzi:2017vha}. For example,   the four-point correlation function (generating function of all four-point correlators of $J$) in the free boson CFT reads \cite{Didenko:2012tv}
\begin{align*}\hspace{-0.7cm}\langle J JJJ\rangle_{F.B.}&=\frac{1}{|x_{12}||x_{23}||x_{34}||x_{41}|}\times \\
&\times \cos(Q^2_{13}+Q^{3}_{24}+Q^4_{31}+Q^1_{43})\cos(P_{12})\cos(P_{23})\cos(P_{34})\cos(P_{41})\\
&+\mbox{permutations}\,.
\end{align*}
Here $Q$ and $P$ are the standard conformally-invariant variables \cite{Giombi:2011rz}. 

\paragraph{Three-dimensional bosonization duality.} The $3d$ bosonization duality is a recently discovered type of duality that takes place in vector models \cite{Giombi:2011kc, Maldacena:2012sf, Aharony:2012nh,Aharony:2015mjs,Karch:2016sxi,Seiberg:2016gmd}. In what follows, we deal with the four simplest vector models. These models are constructed by taking an order parameter, which is an $N$-component complex scalar $\phi^i$ or Dirac fermion $\psi^i$, and writing down the most general $U(N)$-invariant action:
\begin{align*}
    \frac{k}{4\pi}S_{CS}(A)+\text{Matter}
    \begin{cases}
        \displaystyle(D\phi^i)^2 &\text{\stressBlue{free boson}}\\
        \displaystyle (D\phi^i)^2 +g(\phi^i\phi^i)^2 & \text{\stressBlue{Wilson--Fisher (Ising)}}\\
        \displaystyle\bar\psi\slashed D\psi & \text{\stressRed{free fermion}}\\
        \displaystyle \bar\psi\slashed D\psi +g(\bar\psi \psi)^2 & \text{\stressRed{Gross--Neveu}}
    \end{cases}
\end{align*}
where $S_{CS}=\int \Tr[A dA+\tfrac23 A^3]$ is the Chern--Simons action. Other gaugings, e.g. $O(N)$-gauging, are straightforward. There is a huge variety of vector models with different matter multiplets, with and without supersymmetry. Coupling $g$ is not a free parameter, it has to be tuned to the fixed point, so that the model is conformally invariant. The true parameters are the number of fields $N$ and the Chern--Simons level $k$. It is convenient to work with $\lambda=N/k$, which is an analog of the t'Hooft coupling in the large-$N$ limit. Note that due to the Chern--Simons term and covariant derivatives, even the $g=0$ models are not free and it is difficult to say anything concrete about them. 

The $3d$-bosonization duality can be formulated with the help of the following diagram:
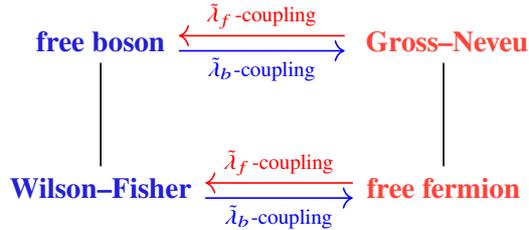
\begin{center}
\begin{tikzcd}[row sep=huge, column sep=huge]
\text{\stressBlue{free boson}}  \arrow[r, "{\text{$\tilde{\lambda}_b$-coupling}}"', shift right=0.1cm, blue] \arrow[r,leftarrow, "{\text{$\tilde{\lambda}_f$-coupling}}", shift left=0.1cm,red] \arrow[d, black,dash]
& \text{ \stressRed{Gross--Neveu}} \arrow[d, black,dash] \\
\text{\stressBlue{Wilson--Fisher}} \arrow[r, blue, "{\text{$\tilde{\lambda}_b$-coupling}}"', shift right=0.1cm] \arrow[r, leftarrow, red, "{\text{$\tilde{\lambda}_f$-coupling}}",shift left=0.1cm]
&  \text{\stressRed{free fermion}}
\end{tikzcd}  
\end{center}
The conjecture is that all gauge-invariant observables should be identical along the double-arrows at the top and at the bottom, i.e. the four theories are dual pairwise. The vertical lines indicate that the 'top' and 'bottom' models are related via RG-flows. The upper line connects the free boson theory to  the Gross--Neveu model, while the bottom line connects the free fermion theory to the Wilson--Fisher model. The red color is for fermionic matter and the blue is for bosonic one. Therefore, the conjecture is that along the upper and along the bottom line there are two equivalent descriptions of the same physics, one in terms of a model with fermionic matter and another one in terms of a model with bosonic matter (provided $\lambda$ and $N$ are mapped appropriately). Let us make this a bit more precise. 

Intuitively, it is clear that in the $k\rightarrow \infty$ limit the Chern--Simons action implies the flatness of $A$, $dA+AA=0$. Therefore, $A$ effectively decouples, the only trace of it being in that we have to consider gauge-invariant observables built out of the matter fields. Note that neither $\phi$ nor $\psi$ are gauge-invariant. The $3d$-bosonization duality, as different from the $2d$ one, does not imply any explicit map between $\phi$ and $\psi$. $O(N)$-models are simpler due to the absence of the spin-one current that can be gauged again. 

There is a number of simple observations that can be made in the large-$N$ limit, irrespective of the value of $\lambda$. The simplest gauge-invariant operators are bilinears
\begin{align}
    J_s& =\phi D\cdots D\phi \,,&& J_s=\bar{\psi} \gamma D\cdots D\psi\,.
\end{align}
They reduce to the usual higher spin currents for free theories once the Chern--Simons term is removed, e.g. via $k=\infty$ or just $A=0$. Therefore, the higher spin symmetry is present in $N=\infty$ limit. It is also clear that all local gauge-invariant operators can be built from the higher spin tensors. For example, the simplest class next to $J_s$ are 'double-trace' operators of type $[JJ]$, see e.g. \cite{Giombi:2016zwa} for explicit formulas. Schematically, 
\begin{align}
    [J_{s_1}J_{s_2}]_s&= \sum \pl\ldots  \pl J_{s_1} \pl\ldots\pl J_{s_2} -\text{traces}\,.
\end{align}
Out of $J_{s_1}$ and  $J_{s_2}$ one can build primary operators of spin $s\geq s_1+s_2$. Obviously, (higher) correlation functions of $J_s$ contain full information about OPE coefficients and anomalous dimensions of all the other local gauge-invariant operators. Therefore, we should {concentrate on learning as much as possible about $J_s$.  }

To the leading order the spectrum of higher spin tensors is identical for both theories on the bottom line and for both theories on the top line. The only difference between the top and bottom models is that the spin-zero operator $J_0$ has conformal dimension $1+\mathcal{O}(N^{-1})$ for the top ones and dimension $2+\mathcal{O}(N^{-1})$ for the bottom ones. That the spectra coincide in the large-$N$ limit for the dual models is the first indication of the duality. 

\paragraph{Slightly-broken higher spin symmetry.} 
The main question is what happens to $J_s$ once we depart from the $N=\infty$ limit. Conserved higher spin tensors signal that the theory is a free one in disguise \cite{Maldacena:2011jn,Boulanger:2013zza,Alba:2013yda,Alba:2015upa}. Therefore, they cannot be conserved. However, the spectrum of vector models is very sparse and there are not so many operators to violate the conservation law: 
\begin{align}\label{noncons}
    \pl \cdot J_s&= \frac{1}{N} \sum_{s_1+s_2\leq s} C^{s}_{s_1,s_2} [J_{s_1} J_{s_2}]_s\,.
\end{align}
Here the structure constants $C^{s}_{s_1,s_2}$ contain the dynamical input, while the form of the double-trace operators is fixed by conformal symmetry to the leading order.\footnote{We note that a simple spin/conformal dimension counting ensures that in the `worst case scenario' one can find in addition $N^{-2}[JJJ]$ on the right-hand side, which happens for the models with $J_0$ of dimension $1+\mathcal{O}(N^{-1})$. }

It was shown in \cite{Maldacena:2012sf}, that the non-conservation law \eqref{noncons} is powerful enough as to fix the form of three-point functions of $J_s$. The idea was to abstract the non-conservation law from any concrete microscopical realization without making any assumptions about structure constants $C^{s}_{s_1,s_2}$. The non-conservation law should be combined with the corresponding violation (or, more precisely, deformation) of the action of charges $Q_s$ on $J_s$:
\begin{align}\label{Qdef}
    [Q_{s_1}, J_{s_2} ]&= \sum_s A_{s_1,s_2}^s J_s + \frac{1}{N}\sum B_{s_1,s_2}^{l_1,l_2,k} [J_{l_1}J_{l_2}]_k\,.
\end{align}
Here $A$'s encode the structure constants of the higher spin algebra and $B$'s signal a violation of the Lie algebra module structure. Now, one can use the fact that three-point functions are fixed up to three numbers per each triplet of spins:
\begin{align}\label{Treep}
     \langle  J_{s_1}J_{s_2}J_{s_3} \rangle\sim a_{s_1,s_2,s_3} \langle J_{s_1}J_{s_2}J_{s_3}\rangle_b+b_{s_1,s_2,s_3}\langle J_{s_1}J_{s_2}J_{s_3}\rangle_f+c_{s_1,s_2,s_3}\langle J_{s_1}J_{s_2}J_{s_3}\rangle_o\,.
\end{align}%
The first two structure can be computed in the free boson and free fermion theory, respectively. The last one, called odd, does not show up in any free theory. Nevertheless, it can be shown to exist and to be unique. Constants $a,b,c$ represent the dynamical information about a given theory and can depend on spins and coupling constants. A remarkable result \cite{Maldacena:2012sf}, is that the slightly-broken higher spin symmetry along can fix all the structure constants in \eqref{noncons},  \eqref{Qdef} and \eqref{Treep} in terms of only two phenomenological parameters $\tilde{N}$ and $\cos^2\theta=1/(1+\tilde{\lambda}^2)$. The final expression for the three-point functions reads\footnote{Note that this requires the building blocks to be properly normalized, so that the spin dependence is trivial. The exact form of all the structures can be found in \cite{Skvortsov:2018uru} in the light-cone gauge and in \cite{Giombi:2016zwa} manifestly Lorentz invariant expressions are given for `almost' all the structures. }
\begin{align}
     \langle  J_{s_1}J_{s_2}J_{s_3} \rangle= \tilde{N}\left [\cos^2 \theta\langle J_{s_1}J_{s_2}J_{s_3}\rangle_b+\sin^2\theta\langle J_{s_1}J_{s_2}J_{s_3}\rangle_f+\cos\theta\sin\theta\langle J_{s_1}J_{s_2}J_{s_3}\rangle_o\right]\,.
\end{align}%
Similar results have been obtained for other three-point functions and some four-point correlators, see e.g.  \cite{Li:2019twz,Kalloor:2019xjb,Turiaci:2018nua,Jain:2021gwa,Silva:2021ece}. This indicates that the slightly-broken higher spin symmetry can be powerful enough to fix all correlation functions! It remains to understand what it is mathematically.

\paragraph{Slightly-broken higher spin symmetry via strong homotopy algebras.} At this point, it is important to understand what "slightly-broken higher spin symmetry" is. The currents are no longer conserved \eqref{noncons}; the action of charges on higher spin currents gets deformed \eqref{Qdef}; the charges themselves cannot form a Lie algebra. Therefore, we cannot abstract a single structure -- higher spin algebra -- and deform it as a Lie algebra. In fact, the algebra does not even have such deformations. 

Since the non-conservation operator is built from the higher spin currents, it should be clear that one cannot disentangle $\hs$ from its action on $J$'s. One has to deform the whole package \cite{Sharapov:2018kjz}: algebra $\hs$ plus its module $J$. The structure that accommodates this is $L_\infty$-algebra. The starting point is an $L_\infty$-algebra $\mathcal{L}=\mathcal{L}_{0}\oplus \mathcal{L}_1$ concentrated in degree zero and one, with $\mathcal{L}_1$ being reserved for $\hs$ and $\mathcal{L}_0$ for the $\hs$-module that $J$'s form:
\begin{align*}
    [\delta_{\xi_1},\delta_{\xi_2}]&= l_2(\xi_1,\xi_2)\,, & \delta_\xi J&=l_2(\xi,J) \,, && \xi\,,\xi_{1,2}\in\hs \,.
\end{align*}
From now on we can abstract $J$'s and think of them as of specific $\hs$-module $\mathcal{L}_0$ rather than concrete operators. What one can do now is to look for the most general deformation 
\begin{align*}
   \delta_\xi J&= l_2(\xi,J)+\,l_3(\xi,J,J)+\ldots\,, & [\delta_{\xi_1},\delta_{\xi_2}]&=\delta_{\xi}\,,  
\end{align*}
where $\xi =l_2(\xi_1,\xi_2)+\,l_3(\xi_1,\xi_2,J)+\ldots$ Self-consistency requires $l_n$'s to form an $L_\infty$-algebra.

There is one specific feature of the problem that allows us to apply the results of section \ref{sec:int}, i.e. to reduce $L_\infty$ to $A_\infty$. Higher spin tensors $J_s$, being bilinear operators, belong to the tensor product $|\phi\rangle \otimes |\phi\rangle $ or $|\psi\rangle \otimes |\psi\rangle$. Higher spin algebra $\hs$ is the endomorphism algebra of one-particle states $|\phi\rangle$ or $|\psi\rangle$. As such it is formally isomorphic to $|\phi\rangle \langle \phi|$ or $|\psi\rangle \langle \psi|$. Recall that the latter two tensor products are isomorphic since the higher spin algebra of free boson and free fermion are identical. Also, $|\phi\rangle \otimes |\phi\rangle $ is formally isomorphic to $|\phi\rangle \langle \phi|$, \textit{idem}. for $\psi$. Therefore, we can map the module of higher spin tensors onto $\hs$. The difference between $|\phi\rangle \otimes |\phi\rangle $ and $|\phi\rangle \langle \phi|$ is by the map $R: |\phi\rangle\rightarrow \langle \phi|$, which is realized via the inversion map $R$. Inversion $R$ is a discrete automorphism of the conformal and, hence, of the higher spin algebra. Finally, the underlying associative structure of $\hs$ is important again and leads to the following realization of the initial data: 
\begin{align*}
    l_2(a,b)&= a\star b-b\star a \,, & l_2(a,v)&= a\star v-v \star R(a) \,, && a,b\in \mathcal{L}_1\,,\quad v\in \mathcal{L}_0\,.
\end{align*}
The last step is to encode  $R$ in the smash product $A_0=\hs \ltimes \mathbb{Z}_2$, where $\mathbb{Z}_2=\{e, R\}$. Now, one can apply the powerful machinery of section \ref{sec:int}, see \cite{Sharapov:2018kjz}. While $\hs$ does not have any deformations as an associative algebra, $A_0$ is soft and can be deformed into $A_u$, with $u$ being a formal parameter.\footnote{For a multi-parameter deformation there is always an overall scale of the parameters that we call $u$ here. It is important to count orders in the expansion, the rest of the parameters, if any, being implicit. } This gives the deformation quantization of the simplest Poisson Orbifold $\mathbb{R}^2/\mathbb{Z}_2$ discussed in section \ref{sec:DQ} (one needs to copies of this Poisson Orbifold since $A_2=A_1 \otimes A_1$). As a result, there is an explicit description of all $l_n$'s. 

In order to discuss the bosonization duality it is also important to understand what is the most general deformation of $l_2$'s that are based on a given $\hs$. Restricting for simplicity to the $O(N)$ case, i.e. even spin currents, it may be shown  \cite{Sharapov:2020quq} that the deformation depends on two phenomenological parameters. They can be related to the microscopical parameters $N$ and $k$, if needed.

\paragraph{Correlation functions as invariants.} For the unbroken higher spin symmetry the correlation functions are given by simple $\hs$-invariants of type $\Tr_\star[\bullet]$, \eqref{correlatorsFree}. It was shown in \cite{Sharapov:2020quq} that these invariants are unique, i.e. there are no other invariants that can serve as correlation functions. Now, one can ask if it is possible to deform them
\begin{align}\label{correlatorsDef}
    \langle J_1\ldots J_n\rangle= \Trace{0}{\wv_1\star \cdots \star \wv_n}+\ldots
\end{align}
in such a way that they remain invariant under the deformed higher spin symmetry, i.e. under the $L_\infty$-transformations. Here we also trivially rewrote $\trA_\star[\bullet]$ on $\hs$ as $\Trace{0}{\bullet}$ on $A_0$, since $\hs\in A_0$ and $A_0$ admits a trace as well. In what follows it is important that the trace can be deformed to a trace $\Trace{u}{\bullet}$ on $A_u$.

It is worth noting that there is a simple generating function of the free CFT correlators
\begin{align}
 W_\text{free}[\wv]&=\mathrm{P.p.}\, \Trace{\star}{\log_\star [1-u^{-1}\wv]}\,,
\end{align}
where $\mathrm {P.p.} $ stands for the principal part of the Laurent series in $u$. Note that $u$ is introduced artificially here and is not present in the $\star$-product. Now, it is possible to show that the free CFT correlators admit a smooth deformation that is invariant under the full $L_\infty$-algebra \cite{Gerasimenko:2021sxj,Sharapov:2022fos}. The final result can be expressed as a very similar generating function:
\begin{align}\label{sbhscorr}
    W_{\text{SBHS}}[\wv]&=\mathrm{P.p.}\,\Trace{u}{\log_\circ[1-u^{-1}\wv]}\,, 
\end{align}
where $u$ is now present in the deformed $\circ$-product and in the trace, cf. \eqref{defproduct}. 

It remains now to compute the corrections to the free CFT correlators. However, several important observations can be made without having to do that. First of all, it can be shown \cite{Sharapov:2020quq} that the deformed invariants -- the $L_\infty$-invariants -- are unique. Therefore, if the idea of the slightly-broken higher spin symmetry is correct, there is a unique $L_\infty$-invariant to serve as a correlation function. This statement proves, in principle, the bosonization duality since the invariant is fixed by the symmetry and does not require any microscopical realization via Chern--Simons vector models, i.e. it has to be the same both for bosonic and fermionic matter. Secondly, a very simple observation is that for any $n$ the $n$-point correlation functions have to have the following form
        \begin{align*}
            \langle J\ldots J\rangle &= \sum \langle \text{fixed} \rangle_i \times \text{params}\,,
        \end{align*}
i.e. it has to be a sum of a finite number of fixed conformally invariant structures multiplied by the powers of the phenomenological deformation parameters. Considering unitarity, the two deformation parameters $u_{1,2}$ have to be of the form $u_1=ue^{+i\theta}$, $u_2=ue^{-i\theta}$. This is consistent with all available correlation functions \cite{Giombi:2011kc, Maldacena:2012sf,Aharony:2012nh, GurAri:2012is,Giombi:2016zwa,Li:2019twz,Kalloor:2019xjb,Turiaci:2018nua,Gandhi:2021gwn,Jain:2020puw,Jain:2021gwa,Silva:2021ece}. It is intriguing that expression \eqref{sbhscorr} bears close similarity to the partition function of a one-loop exact theory.        
        
\section*{Acknowledgments}
\label{sec:Aknowledgements}
The work of E.S. was partially supported by the European Research Council (ERC) under the European Union’s Horizon 2020 research and innovation programme (grant agreement No 101002551) and by the Fonds de la Recherche Scientifique --- FNRS under Grant No. F.4544.21. A. Sh. gratefully acknowledges the financial support of the Foundation for the Advancement of Theoretical Physics and Mathematics “BASIS”. The results on the deformation of $A_\infty$-algebras were obtained with an exclusive support of the Ministry of Science and Higher Education of the Russian Federation (Project No. 0721-2020-0033).

\setlength{\bibsep}{0.0pt}
\footnotesize
\providecommand{\href}[2]{#2}\begingroup\raggedright\endgroup

\end{document}